\documentclass{article}
\usepackage{smc2019}
\usepackage{times}
\usepackage{ifpdf}
\usepackage[english]{babel}
\usepackage{cite}

\usepackage{amsmath} 
\usepackage{booktabs} 

\def\papertitle{Learning to Generate Music with BachProp}
\def\firstauthor{Florian Colombo}
\def\secondauthor{Johanni Brea}
\def\thirdauthor{Wulfram Gerstner}

\newif\ifpdf
\ifx\pdfoutput\relax
\else
   \ifcase\pdfoutput
      \pdffalse
   \else
      \pdftrue
\fi

\ifpdf 
  \usepackage[pdftex,
    pdftitle={\papertitle},
    pdfauthor={\firstauthor, \secondauthor, \thirdauthor},
    bookmarksnumbered, 
    pdfstartview=XYZ 
   ]{hyperref}

  \usepackage[pdftex]{graphicx}
  \graphicspath{{./figures/}}
  \DeclareGraphicsExtensions{.pdf,.jpeg,.png}

  \usepackage[figure,table]{hypcap}

\else 
  \usepackage[dvips,
    bookmarksnumbered, 
    pdfstartview=XYZ 
  ]{hyperref}  

  \usepackage[dvips]{epsfig,graphicx}
  \graphicspath{{./figures/}}
  \DeclareGraphicsExtensions{.eps}

  \usepackage[figure,table]{hypcap}
\fi

\hypersetup{
    colorlinks,%
    citecolor=black,%
    filecolor=black,%
    linkcolor=black,%
    urlcolor=black
}

\title{\papertitle}

 \threeauthors
   {\firstauthor} {School of Computer Science\\ and School of Life Sciences \\ 
   \'Ecole Polytechnique F\'ed\'erale\\ de Lausanne, Switzerland\\%
     {\tt \href{mailto:florian.colombo@epfl.ch}{florian.colombo@epfl.ch}}}
   {\secondauthor} {School of Computer Science\\ and School of Life Sciences \\ 
   \'Ecole Polytechnique F\'ed\'erale\\ de Lausanne, Switzerland \\ %
     {\tt \href{mailto:johanni.brea@epfl.ch}{johanni.brea@epfl.ch}}}
   {\thirdauthor} { School of Computer Science\\ and School of Life Sciences \\ 
   \'Ecole Polytechnique F\'ed\'erale\\ de Lausanne, Switzerland\\ %
     {\tt \href{mailto:wulfram.gerstner@epfl.ch}{wulfram.gerstner@epfl.ch}}}

\begin{document}
\capstartfalse
\maketitle
\capstarttrue

\begin{abstract}
As deep learning advances, algorithms of music composition increase in performance. 
However, most of the successful models are designed for specific musical structures. 
Here, we present BachProp, an algorithmic composer that can generate music scores in many styles given sufficient training data. 
To adapt BachProp to a broad range of musical styles, we propose a novel representation of music and train a deep network to predict the note transition probabilities of a given music corpus. 
In this paper, new music scores generated by BachProp are compared with the original corpora as well as with different network architectures and other related models. A set of comparative measures is used to demonstrate that BachProp captures important features of the original datasets better than other models and invite the reader to a qualitative comparison on a large collection of generated songs.
\end{abstract}

\section{Introduction}
\label{introduction}



In search of the computational creativity frontier \cite{colton2012computational}, machine learning algorithms are more and more present in creative domains such as painting \cite{mordvintsev2015inceptionism,gatys2016image} and music \cite{sturm2016music, colombo2017deep, hadjeres2016deepbach}. 
Already in 1847, Ada Lovelace predicted the potential of analytical engines for algorithmic music composition \cite{lovelace1843notes}. 
Current models of music generation include rule based approaches, genetic algorithms, Markov models or more recently artificial neural networks \cite{fernandez2013ai}.

One of the first artificial neural networks applied to music composition was a recurrent neural network  trained to generate monophonic melodies \cite{todd1989connectionist}. 
In 2002, networks of long short-term memory (LSTM) \cite{hochreiter1997long} were applied for the first time to music composition, so as to generate Blues monophonic melodies constrained on chord progressions \cite{eck2002finding}. 
Since then, music composition algorithms employing LSTM units, have been used to generate monophonic \cite{sturm2016music, colombo2017deep} and polyphonic music \cite{BL2012,Lattner2018,bachbotismir,hadjeres2016deepbach} or to harmonize chorales in the style of Bach  \cite{bachbotismir,hadjeres2016deepbach}. 
However, most of these algorithms make strong assumptions about the structure of the music they model. 

Here, we present a neural composer algorithm named \emph{BachProp} designed to generate new music scores in an arbitrary style implicitly defined by the corpus of training data. 
To this end, we do not assume any specific musical structure of the data except that it is composed of sequences of notes that are characterized by pitch, duration and time-shift relative to the previous note. 

In the following, we start by contrasting our representation of music to previous propositions \cite{BL2012,bachbotismir,hadjeres2016deepbach,magenta2016polyphonyrnn} with a focus towards training style-agnostic generative models of music.
We then introduce our algorithm and compare BachProp with other models on a standard datasets of chor- ales written by Johann Sebastian Bach \cite{BachChorales} and establish new benchmarks on the musically complex datasets of MI- DI recordings by John Sankey \cite{JohnSankey} and string quartets by Haydn and Mozart \cite{StringQuartets}.
Finally, as the evaluation and comparison of generative models is not trivial \cite{Theis2015}, we invite the reader, first, to a subjective comparison on a large collection of samples generated from the different models on the accompanying media webpage\cite{media} and, second, we propose a new set of metrics to quantify differences between the models.
Preliminary versions of our work have been made available on arXiv \cite{colombo2018bachprop, colombo2018learning}.

\section{Related work}
\label{sec:repr}

Unlike approaches to image generation, where the standard data consists of rows and columns of pixel values for multiple color channels, approaches to music generation lack a standard representation of music data.
This is reflected by the zoo of music notation file formats (ABC, LilyPond, MusicXML, NIFF, MIDI) and the fact that lossless conversion from one to the other is usually not possible.
The MIDI file format captures most features of music, like polyphony, dynamics, micro tuning, expressive timing and tempo changes.
But its representational richness and the possibility to represent the exact same song in multiple ways, make it challenging to work directly with MIDI.
Therefore, all approaches discussed in the following use a first preprocessing step to transform all songs into a simpler representation.
The subsequent design choices of the generative model are heavily influenced by this first preprocessing step.

DeepBach \cite{hadjeres2016deepbach} is designed exclusively for songs with a constant number of voices (e.g. four voices for a typical Bach chorale) and a discretization of the rhythm into multiples of a base unit, e.g. 16\textsuperscript{th} notes. 
The model achieves good results not only in generating novel songs but allows also in reharmonizing given melodies while respecting user-provided meta-information like the temporal position of fermatas.
The model works with a Gibbs-sampling-like procedure, where, for each voice and time step, one note is sampled from conditional distributions parameterized by deep neural networks.
The conditioning is on the other voices in a time window surrounding the current time-step. 
Additionally a ``temporal backbone'' signals the position of the current 16\textsuperscript{th} note relative to quarter notes and other meta-information.
A special hold symbol can also be sampled instead of a note, to represent notes with a duration longer than one time-step.

BachBot \cite{bachbotismir} and its Magenta implementation Polyphony-RNN \cite{magenta2016polyphonyrnn} contain no assumption about the number of voices; they can be fit to any corpus of polyphonic music, if the rhythm can be discretized into multiples of a base unit, e.g. 16\textsuperscript{th} notes.
Songs are represented as sequences of {\sc new\_note(pitch), cont\_note(pitch)} and {\sc step\_end} events, where the {\sc step\_end} event indicates the end of the current time-step. 
Between two {\sc step\_end} events, typically several {\sc new\_note(pitch)} and {\sc cont\_note(pitch)} events can be found sorted by {\sc pitch}.
A generative model parametrized by a recurrent neural network model is fit to these sequences of events, in the same way as recurrent neural network models are used for language modeling on a character- or word-level \cite{Sutskever2011,Graves2013,Mikolov2012}.

%
Common to the models discussed above is a discretization of time into multiples of a base unit like the 16\textsuperscript{th} note.
This limits the representable rhythms considerably; e.g. triplets, grace notes or expressive variations in timing cannot be represented in this way.
To overcome this limitation, \cite{Oore2018} replace the repertoire of symbols employed by the Polyphony-RNN by {\sc note\_on, note\_off, time\_shift} and {\sc set\_velocity} events, where the {\sc time} {\sc\_shift} events allows the model to move forward in time by multiples of 8 ms up to 1 second and the {\sc set\_velocity} events allow to model the loudness of a note (which depends on the piano on the velocity with which a key is pressed).


\section{Method}

\newcommand{\note}{\mathbf{note}}
\newcommand{\song}{\mathbf{song}}
\newcommand{\prob}{Pr}

In written music, the $n$\textsuperscript{th} note $\note[n]$ of a piece of music $\song = (\note[1], \ldots, \note[N])$ can be characterized by its pitch $P[n]$, duration $T[n]$  and the time-shift $dT[n]$ of its onset relative to the previous note, i.e. $\note[n] = (dT[n], T[n], P[n])$.
The time-shift $dT[n]$ is zero for notes played at the same time as the previous note. 
In contrast to most other approaches that discretize the time into multiples of a base unit (except e.g. \cite{Oore2018}), we round all durations into a set of defined musical durations which allows a more faithful representation of timing that is limited only by the number of possible values considered for $T[n]$ and $dT[n]$. 
For example, our representation allows to easily and without any distortion represent 32\textsuperscript{nd} notes, triplets and double dotted notes in the same dataset. 
As well as any other more complex note durations that can be needed for specific corpora. 
To achieve that, each duration in the original corpus is mapped to the closest duration in a set of integer multiples of atomic durations. 
For this work, we considered 64\textsuperscript{th} and 32\textsuperscript{nd} triplets as atomic durations.


Our approach is to approximate probability distributions over note sequences in music scores $\song_1,\ldots, \song_S$ with distributions parameterized by recurrent neural networks and move its weights $\theta$ towards the maximum likelihood estimate
\begin{equation}
    \theta^* = \arg\max_\theta \prob(\song_1,\ldots, \song_S|\theta) \, ,
\end{equation} 
Since each note in each song consists of the triplet ($dT, T, P$) we can parametrize the distributions in a similar way as the pixel-RNN \cite{vandenOord2016} that was developed for the (red, green, blue) triplets of pixels in images. 
Importantly, our model takes into account that pitch and duration of a note are generally not independent. For example in classical music, the fundamental, e.g. the note C in a piece written in C major, tends to be longer than other notes. 

In the following we describe in more details our representation of music, the structure of the model and our approach to comparing different models that use different representations of music.

\subsection{Conversion of MIDI files into our representation of music}
\label{sec:midi}


%

\begin{figure}[h]
\centering
\includegraphics[width=.45\textwidth]{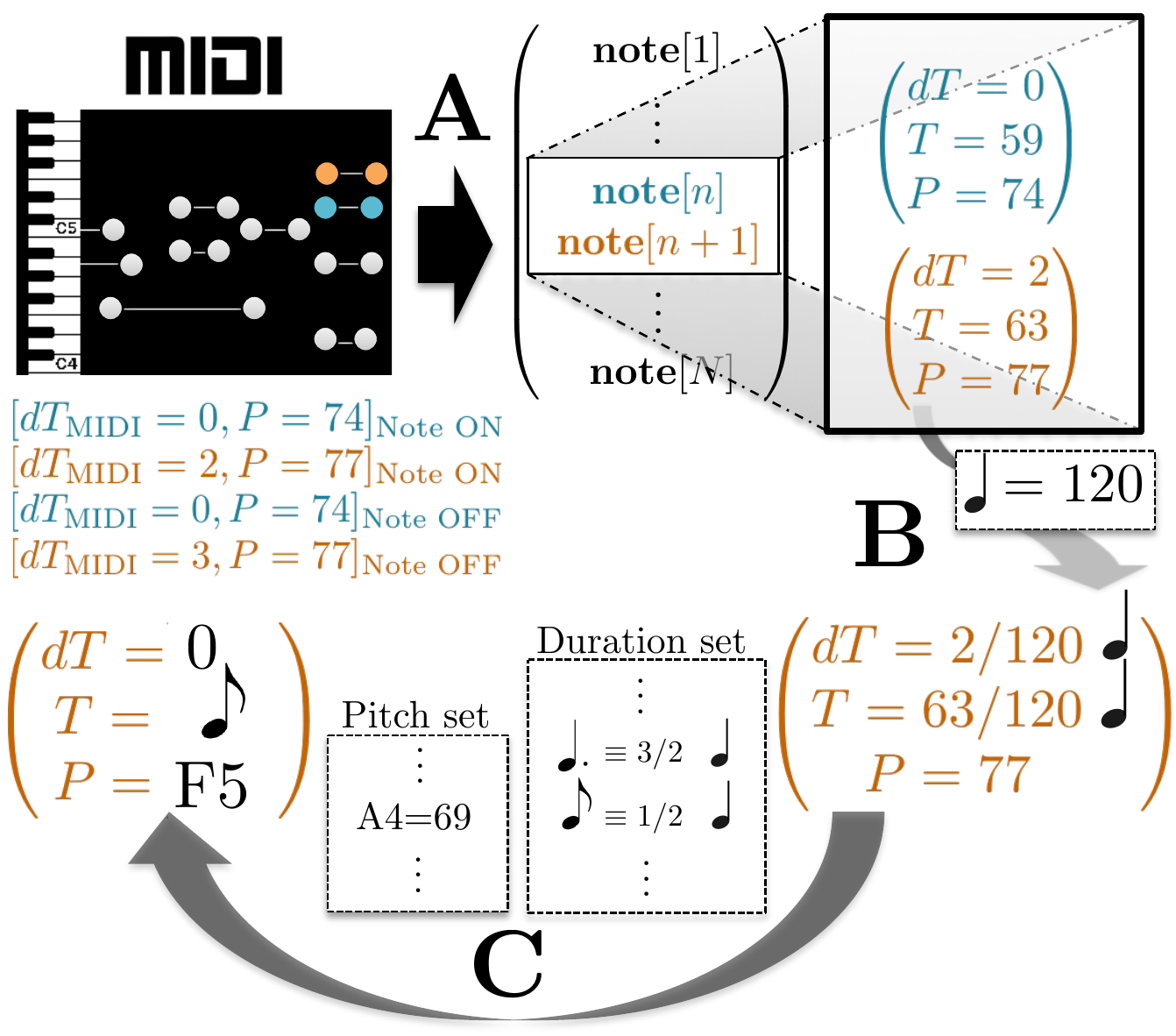}
\caption{\small \textbf{From MIDI to our representation of music.} 
An illustration of the steps involved in the proposed conversion of MIDI sequences. See text for details.
}
\label{fig:midi}
\end{figure}

%

%

A MIDI file contains a header (meta parameters) and possibly multiple tracks that contain a sequence of MIDI messages. 
For BachProp, we merge all tracks and consider only the MIDI messages defining when a note starts ({\sc on} events) or ends ({\sc off} events). 
For each  {\sc on} event we look forward at the next {\sc off} event with the same pitch $P$ to convert sequences of MIDI messages into a sequences of notes (Figure \ref{fig:midi}{\bf A}). We then translate timings from the internal MIDI {\sc tick} representation to quarter note lengths (\autoref{fig:midi}{\bf B}).

Next, we round all durations $T[n]$ such that they are in the set of possible note lengths (duration set in \autoref{fig:midi}{\bf C}) expressed in units of a quarter note, similar to durations in standard music notation software.
Similarly, we round the time-shifts $dT[n]$ to 0 or one of the possible note lengths.
Mapping to the closest value in the set removes temporal jitter around the standard note duration that may have been introduced accidentally at the moment of recording the MIDI file (\autoref{fig:midi}{\bf C}).
While this standardization may be desired when expressive timing is not taken into account, it is straightforward to extend the duration dictionary to include also values that allow to model expressive timing. 
However, we believe that a good generative model of music should be trained to model the music structure only. 
If the music representation is introducing additional temporal dependencies, the model will need to learn these interfering structures as well. 
In that sense, the processing we designed for BachProp and presented in this section allows it to focus on the essential structure of music. 


In order for BachProp to learn tonality, during training and before each new epoch, we randomly transpose every song within the available bounds of the pitch set. 
For each song we compute one of the possible shift of semitones and apply it as an offset to all pitches within the song. 
Because a single MIDI sequence will be transposed with up to 20 offsets, this augmentation method allows BachProp to learn the temporal structure of music on more examples. 

Finally, we add an artificial note at the beginning and end of each score. 
After training, the inaudible `boundary note' is used by the model to seed and end the generation of songs.

\begin{figure}
\centering
\includegraphics[width=.45\textwidth]{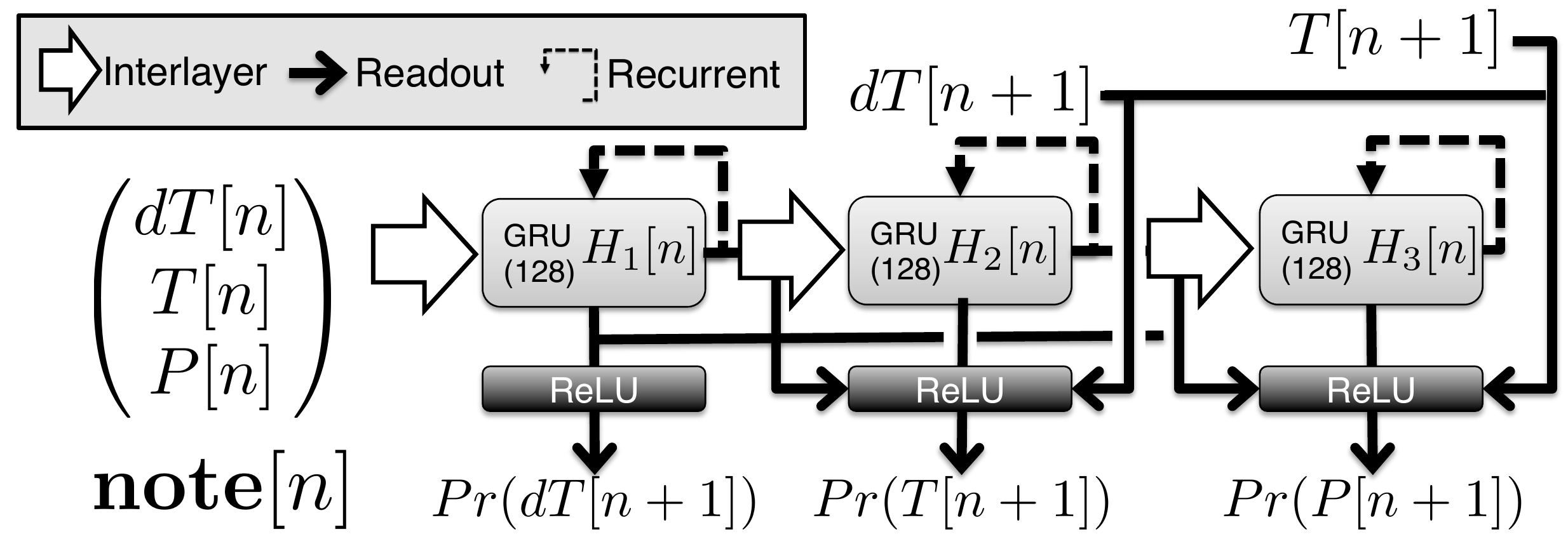}
\caption{\small \textbf{BachProp neural architecture.} See text for details.
}
\label{fig:architecture}
\end{figure}

\subsection{The BachProp neural network}
\label{model}



We employ a deep GRU \cite{chung2014empirical} network with three consecutive layers as schematized in Figure \ref{fig:architecture}. 
The network's task is to infer the probability distribution over the next possible notes from the representation of the current note and the network's internal state (the network representation of the history of notes). 



%
The probability of a sequence of $N$ notes $\note[1:N] = (\note[1], \ldots, \note[N])$ is given by
\begin{align}
    Pr(\note[1:N]) = \nonumber\\Pr(\note[1])\prod_{n=1}^{N-1}Pr(\note[n+1]|\note[1:n])\, .
 \label{eq:prS}
\end{align}
Each term on the right hand side can be further split into 
\begin{align}
Pr(\mathbf{note}[n+1]|\mathbf{note}[1:n]) =\nonumber\\
Pr(dT[n+1]|\mathbf{note}[1:n]) \times\nonumber\\ 
Pr(T[n+1]|\mathbf{note}[1:n],dT[n+1])\times\nonumber\\ 
Pr(P[n+1]|\mathbf{note}[1:n],dT[n+1],T[n+1])\, .
\label{eq:prE}
\end{align}


The goal of training the Bachprop network with parameters $\theta$ is to approximate the conditional probability distributions on the right hand side of Equation (\ref{eq:prE}).

In the BachProp network (\autoref{fig:architecture}), the conditioning on the history $\note[1:n]$ is implemented by the values of the shared hidden states.
The hidden state is composed of 3 recurrent layers with 128 gated-recurrent units (GRU).
The state $H_1[n]$ of the first hidden layer is updated with input $\note[n]$ and previous state $H_1[n-1]$. The state of the upper layers $H_i[n]$ for $i=2, 3$ are updated with inputs $H_{i-1}[n]$ and $H_i[n-1]$.

$\note[n]$ is represented by three one-hot vectors encoding separately $dT[n]$, $T[n]$ and $P[n]$, i.e. every entry in these vectors is 0 but the one mapped to the value $x[n]$ for $x=dT, T, P$.
The length of each of these vectors is defined by the size of the respective dictionary $L_x$.
Therefore, each song is encoded as a set of three 2-dimensional binary matrices of size $N_s \times L_x$, where $N_s$ stands for number of notes in song $s$ and $L_x$ for the size of the set of unique time-shifts ($x=dT$), note durations ($x=T$) or keys ($x=P$) present in the original corpus. 
These matrices are zero-padded to account for variable input lengths $N_s$.

During training, songs are presented to the network in sequences of batches containing binary tensors with dimension $B \times N \times L$.
$B=32$ is the number of songs presented together over which the gradient of the error signal is averaged.
$N=128$ is the number of consecutive notes over which the gradients are computed, i.e. we used truncated backpropagation through time. 
$L = L_{dT} + L_T + L_P$ is the size of the entire input vector encoding for $\note[n]$.
Though the gradient is truncated, the recurrent units are stateful: they maintain their states across consecutive batches while the same sets of $B$ songs are being presented. 
When a new batch of $B$ songs is being presented to the network, hidden states are reset.
 
To generate $\note[n+1]$, one third ($H_1[n]$ in \autoref{fig:architecture}) of the full hidden state is fed into a feedforward network with one layer of Rectified linear (ReLU) units and one output softmax layer that represents $Pr(dT[n+1]|H_1[n])\approx Pr(dT[n+1]|\note[1:n])$.
The chosen $dT[n+1]$ together with $H_1[n]$ and $H_2[n]$ is fed into a second feedforward network with one layer of ReLU units and an output softmax layer that represents $Pr(T[n+1]|H_1[n], H_2[n], dT[n+1])\approx Pr(T[n+1]|\note[1:n], dT[n+1])$.
In a similarly way, the pitch is sampled from $Pr(P[n+1]|H_1[n],$ $H_2[n], H_3[n], dT[n+1], T[n+1])\approx Pr(T[n+1]|\note[1:n], dT[n+1], T[n+1])$.  The ReLU and softmax readout layers have the same size $L_x$ as the dictionary of the feature they model.
These three small steps of sampling $dT[n+1]$, $T[n+1]$ and $P[n+1]$ form together one big step from note $n$ to note $n+1$.



The resulting sequence of notes is a newly generated score sampled from BachProp. 
Note that, the temperature of sampling can be adapted to the confidence we give to the model predictions \cite{karpathy2016unreasonable, colombo2017deep}. 
In particular, any model trained with a corpus that exhibits many repetition of patterns, will generate scores with more examples of these repetitions for lower sampling temperatures. Indeed, a lower temperature will reduce the probability to select an undesired note that is not part of the pattern to be repeated. 
Finally, the generated sequence of notes in our representation can easily be translated back to a MIDI sequence by reversing the method schematized in Figure \ref{fig:midi}.

BachProp has been implemented in Python using the Keras API \cite{chollet2015keras}. 
Code is available on GitHub\footnote{\url{https://github.com/FlorianColombo/BachProp}}.

\subsection{Comparison against plagiarism and other models}
Even in well-established domains such as computer vision and image generation, it is not clear how to compare generative models \cite{Theis2015}.
But in order to turn generative models of music eventually into useful tools for composers, they should be able to generate (1) plagiarism-free music of (2) a predefined style or mood that is (3) pleasant to listen to.

A  way of measuring plagiarism is to control overfitting by comparing the loss on training and validation data. 
While this is a simple method it is rather coarse since it works on songs as a whole.
Instead we propose \emph{novelty profiles} that compare the co-occurrence of short note sequences across different data sets. 
A crucial parameter of novelty profiles is the length of a note sequence on which the comparison takes place. 
We adapted the novelty profile, a measure of similarity between any given score and a reference corpus, from \cite{colombo2017deep}. 
For a pattern size of 6 notes, a novelty score of 1 indicates that all patterns of 6 consecutive notes are not present in the reference corpus. 
On the other hand, a note sequence that contains only patterns found in the reference corpus would exhibit a novelty score of 0. 
We define the binary novelty of a single pattern by checking if all three features ($dT[n-m:n]$, $T[n-m:n]$, $P[n-m:n]$) of the notes included in the pattern are found in the same order anywhere in the reference corpus. 
The novelty score of an entire song is the average binary novelty over all possible patterns. 


Models that are trained on the same representation of music can be compared by their likelihood to assess how well they generate pieces of a predefined type.
But if the models represent probability distributions over different spaces, which is quickly the case when different representations are used, they are unfortunately not comparable in terms of likelihood. 
For example, the event based representation from \cite{Oore2018} can in principle produce all possible note sequences. 
But it could also generate nonsensical sequences of multiple consecutive {\sc note\_off} events, without corresponding previous {\sc note\_on} events. 
To nevertheless compare models that build on different representations of music we propose simple statistics like interval distributions that can be applied to the samples of each generative model of music. 

Finally, to compare the pleasantness of the generated music, one can ask people to rate different pieces; an approach that is followed in previous works (e.g. \cite{hadjeres2016deepbach}). 
We also invite the reader to listen to the large collections of non-cherry-picked generated examples \cite{media}.

\section{Results and discussion}
\label{sec:results}


\subsection{Datasets}

We consider four MIDI corpora with different musical structures and styles (see Table \ref{table:results2}). 
The Nottingham database \cite{Nottingham}
contains British and American folk tunes. 
The musical structure of all songs is very similar with a melody on top of simple chords. The Chorales corpus \cite{BachChorales}
includes hundreds of four-part chorales harmonized by Bach. All chorales share some common structures, such as the number of voices and rhythmical patterns. 
For comparison we used the same filtering of songs as DeepBach \cite{hadjeresmaster} to exclude chorales with number of voices unequal four.
We consider both Nottingham and Chorales corpora as homogeneous data sets. The John Sankey data set \cite{JohnSankey}
is a collection of MIDI sequences recorded by John Sankey on a digital keyboard. Even though all songs were composed by Bach, the pieces are rather different. In addition, this data set was recorded live from the digital keyboard and thus we applied the temporal normalization described above. At last, the string quartets data set \cite{StringQuartets}
includes string quartets from Haydn and Mozart. Here again, there is a large heterogeneity of pieces across the corpus. 

Renderings of original and generated scores are available for listening on the webpage containing media for this paper\footnote{\label{media}Media webpage: \url{https://goo.gl/Z4AfPg}} \cite{media}. 
To train BachProp on the different corpora, we used the same network architecture, number of neurons, initialization and learning parameters, but each of the network was trained on a different corpus.

\subsection{Alternative models}

In addition to BachProp, we trained six other models; three BachProp variants to assess the impact of our design choices, one baseline model and two previously published and available artificial composers. 
PolyDAC and IndepBP are direct BachProp variants. 
MidiBP is a version of BachProp that utilizes a different representation of MIDI note sequences inspired by \cite{Oore2018}. 
The two state-of-the-art artificial composers, DeepBach \cite{hadjeres2016deepbach} and PolyRNN \cite{magenta2016polyphonyrnn} allow us to compare scores generated by models of our design with other algorithms. 
The 6th model is a multi-layer perceptron model (MLP) and serves as a baseline control.

{\bf PolyDAC} is a polyphonic version of \cite{colombo2017deep}. 
It models the same conditional distribution as BachProp but instead of reading out the probabilities from shared hidden layer states, it models each note feature with three independent neural networks. 
The time-shift, duration, and pitch networks are composed of three recurrent layers with 16, 128, and 256 GRUs respectively. 
{\bf IndepBP} assumes that all note features are independent from each others. 
As such, $Pr(dT[n+1])$, $Pr(T[n+1])$, and $Pr(P[n+1])$ are read out by three softmax output layers directly from the hidden state of three hidden layers composed of 128 GRUs that takes as input the one-hot encoding of the $n^{\text th}$ note. 
{\bf MidiBP} neural architecture consists of three recurrent layers composed of 128 GRUs. Here, the MIDI note sequences are represented differently. 
While the normalization and  preprocessing is done as described above (\autoref{fig:midi}), we then convert the normalized music score back to the MIDI-like format proposed  in \cite{Oore2018} where in each time step a single on-hot vector defines either a {\sc note\_on} event and its corresponding pitch, a {\sc note\_off} event and its corresponding pitch, or a time-shift and its corresponding duration (defined by our duration representation). Therefore, a single softmax read out layer is used to sample the upcoming MIDI event.
{\bf MLP} has no recurrent layers but 3 feedforward hidden layers of 124 ReLUs each that gets as input the 5 most recent notes $\note[n-4:n]$ together with the current time-shift $dT[n+1]$ and duration $T[n+1]$ to sample the pitch $P[n+1]$. 
To sample the duration $T[n+1]$ and the time-shift $dT[n+1]$, appropriate parts of the input are masked with zeros.

Models BachProp, PolyDAC, MidiBP, IndepBP were trained with truncated back propagation through time and the Adam optimizer \cite{kingma2014adam}. 
The MLP model was trained with standard back propagation and the Adam optimizer.
The mini-batch size is 32 scores, the validation set a 0.1 fraction of the original corpus, and one training epoch consists of updating the network parameters with all training examples and evaluating the performances on the entire validation set. 
Training is stopped when the performances on the validation set saturates and the model leading to the highest accuracy is used for generating new music scores. DeepBach was trained for 15 epochs with the standard settings of the current master branch \cite{hadjeresmaster}. PolyRNN was trained for 26000 steps with the standard settings of the current master branch \cite{magenta2016polyphonyrnn}.

\begin{table}[h]
\label{table:results}
\begin{center}
\begin{small}
\centering
\begin{sc}
\begin{tabular}{l|cccc}
\toprule
Model &NLL 	& $dT$ 	& $T$	& $P$  \\
\midrule
BachProp & 0.419	& 0.97	& 0.91 	& 0.77 \\
PolyDAC	  & 0.647	& 0.97	& 0.94 	& 0.69\\
IndepBP	 & 0.647	& 0.97	& 0.75 	& 0.63\\
MLP	& 0.796	& 0.95	& 0.76 	& 0.49\\

\bottomrule
\end{tabular}
\end{sc}
\end{small}
\end{center}
 \caption{\small {\bf Comparison of architectures on our representation of music.} NLL stands for negative log-likelihood on the validation set. Columns $dT$, $T$ and $P$ indicate the accuracy (fraction of correct predictions) for time-shifts, durations and pitches, respectively.}

\end{table}


\subsection{BachProp performs better than alternative models with same representation}
On the Bach Chorales we find that the BachProp architecture performs considerably better than the alternative architectures using the same representation of music (see \autoref{table:results}). 
As expected, the standard feedforward MLP with ReLUs yields the worst performance. 
It lacks the ability to model long range dependencies, which the other models can do through their recurrent connections. 
When we remove the conditioning on each of probability terms on the right side of Equation (\ref{eq:prE}), as done for the IndepBP model, we get poorer performances. We further observe that sharing a common hidden state allowed BachProp to outperform PolyDAC on the pitch predictions.

\subsection{BachProp performs at least as good as alternatives with different representation}

\def\imagetop#1{\vtop{\null\hbox{#1}}}
\begin{figure}
	\centering
    \begin{tabular}{l}
        {\bf A} \\
        \imagetop{\includegraphics[width=.45\textwidth]{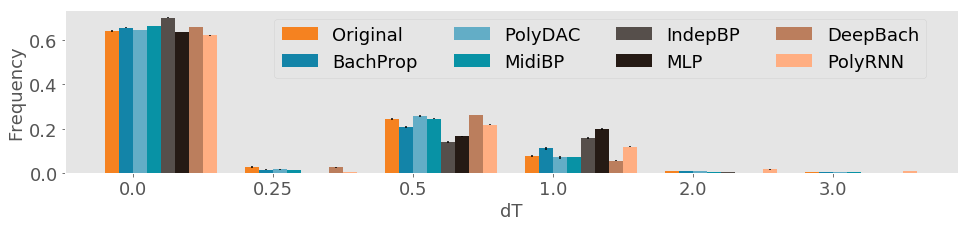}}\\
    	{\bf B} \\
   	\imagetop{\includegraphics[width=.45\textwidth]{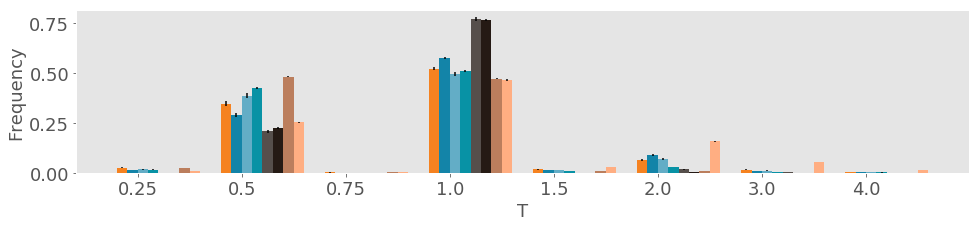}}\\
    	{\bf  C} \\
    	\imagetop{\includegraphics[width=.45\textwidth]{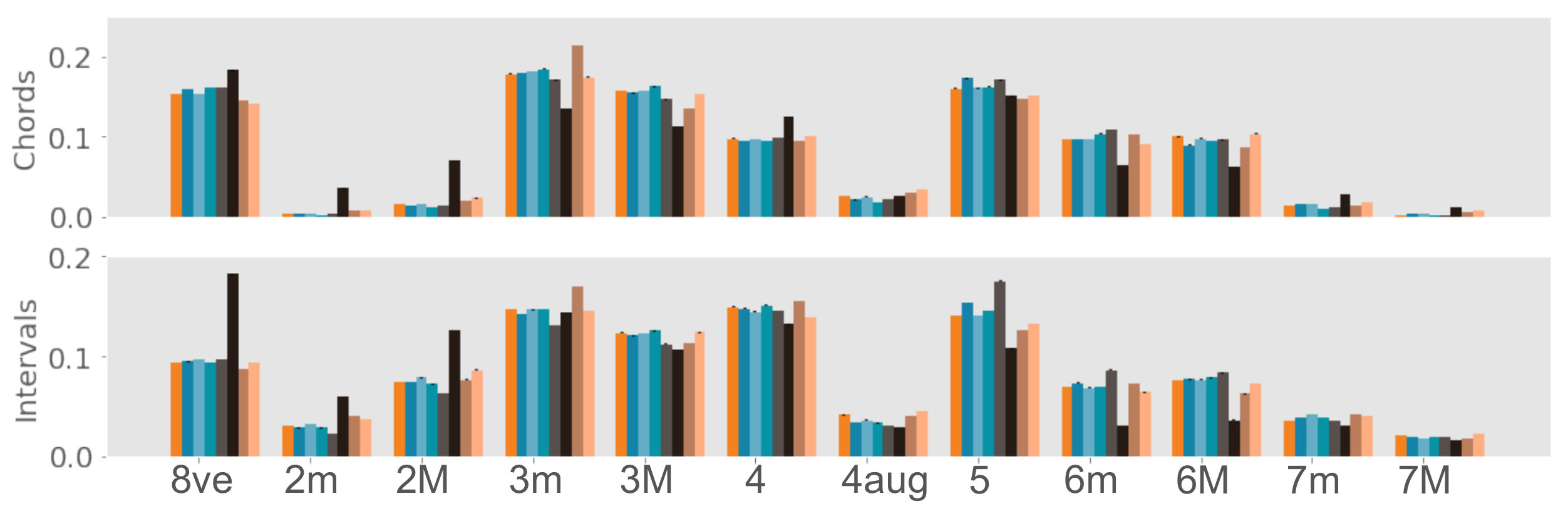}}
    \end{tabular}
	\caption{\small \textbf{Local statistics.} {\bf A} Distribution of $dT$. {\bf B} Distribution of $T$. {\bf C} Distribution of intervals in chords (top) and between each note (bottom). For all figures, we show the mean and standard deviation (in black) obtained with bootstrapping (50\% of the entire corpus resampled 10 times). All models were trained on the Bach Chorales corpus.}
	\label{fig:local}
\end{figure}

To compare models that use a different representation of music, we look at a set of metrics that includes local statistics, song-length statistics and novelty profiles.
To evaluate these metrics for each model, we generated from each model a set containing as many scores as the original Bach Chorales corpus. We include the baseline models from the last section for comparison reasons. 

\subsubsection{Local statistics}
\label{local}

A model that has captured the underlying structure of the sequences of notes present in a corpus, should be able to generate new scores matching the local statistics of what they modeled. 
As such, we suggest to compute the distributions of generated $dT$ and $T$ and compare them to the original corpus distributions as a first metric to evaluate generative models of music. 
Note that for such direct local statistics, a simple n-gram model would match the original distributions perfectly. 
Figure \ref{fig:local}{\bf A} and {\bf B} shows that BachProp and PolyDAC match the original distributions best, followed by MidiBP, DeepBach and PolyRNN, while IndepBP and MLP match the least.

Next, we look at interval distributions. 
An interval is the number of half-tone separating two notes. 
Here, BachProp, PolyDAC, MidiBP and PolyRNN match the distribution quite well. 
DeepBach seems to generate minor thirds considerably more often than present in the training data (Figure \ref{fig:local}{\bf C}).

\subsubsection{Distribution of song lengths}
\label{sec:broad}

\begin{figure}
	\centering
    \begin{tabular}{l}
        {\bf  A} \\
        \imagetop{\includegraphics[width=.45\textwidth]{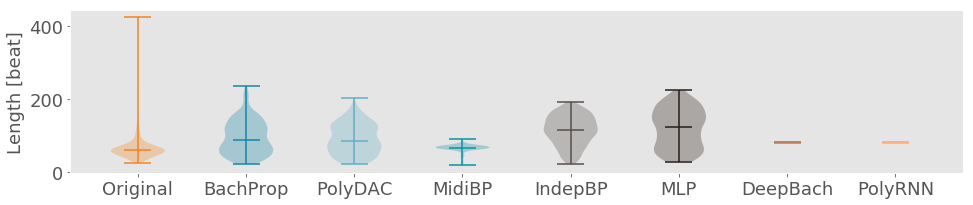}}\\
    {\bf  B} \\
    \imagetop{\includegraphics[width=.45\textwidth]{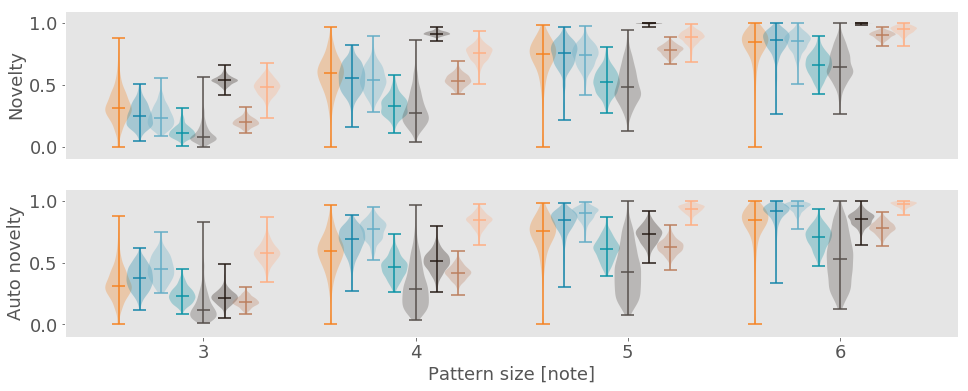}}\\
    \end{tabular}
	\caption{\small \textbf{Song lengths and novelty profiles. } {\bf A} Distribution of the duration of scores in quarter note length. {\bf B} Novelty profile of all corpora with respect to the auto-novelty of the original corpus (top). The auto-novelty profiles of all corpora (bottom). See text for details.
    }\label{fig:broad}
	
\end{figure}

The distribution of song lengths can indicate whether a model captured really long-range dependencies in the training set. 
On this measure MidiBP matches the distribution slightly better than BachProp, PolyDAC, IndepBP and MLP (see \autoref{fig:broad}{\bf A}).
Since DeepBach and PolyRNN do not model score endings, we manually set their duration. 

\subsubsection{Novelty profiles}

In \autoref{fig:broad}{\bf B} (top), we compare the novelty profiles for all models with respect to the original Chorales corpus with which each model was trained. 
We compare the different profiles with the auto-novelty of the reference corpus. 
The auto-novelty is the novelty profile for each song in the reference corpus with respect to the same corpus without the song for which the novelty score is computed. 
It reflects, how similar is the music within the original corpus and is consequently the distribution to match for an ideal generative model of music. 
Here, the only model that is clearly outside the target distribution is the MLP model. 
While the IndepBP and MidiBP models match the target distributions, their novelty distributions for bigger pattern sizes is lower than the original corpus auto-novelty. 
This is an indicator that these models are generating music examples that are too similar to the original data. 
In other words, these models adopted a strategy closer to reproducing or recombining observed patterns rather than inferring the actual temporal dependencies between music notes. 
DeepBach, BachProp and PolyDAC have their medians close and above the original distributions. 
However, DeepBach and PolyRNN have a surprisingly low variance for each of the pattern sizes. 

In \autoref{fig:broad}{\bf B} (bottom) we compare the auto-novelty of all generated corpora with the original corpus. 
An auto-novelty profile exhibiting distributions with lower novelty scores than the original data set, is suspected to generate new music scores of little diversity. 
The auto-novelty profile of BachProp and PolyDAC match the one of the original corpus best.

\subsection{BachProp generates pleasant examples on more complex datasets}
\begin{table}
    
\label{table:results2}
\begin{tiny}
\centering
\begin{sc}
\begin{tabular}{l|cccclc}

\toprule
Dataset &NLL 	& $dT$ 	& $T$	& $P$ & Size [score | note] \\
\midrule
Chorales  &  0.419	& 0.97 	& 0.91	& 0.77 &  357 | 95'337\\
Nottingham  & 0.587 &  0.98 & 0.89 & 0.70 & 1037 | 313'975 \\
John Sankey & 1.002 & 0.89 & 0.77 & 0.45 & 135 | 358'211\\
String quartets  & 0.936 & 0.88 & 0.83 & 0.49 & 215 | 738'739 \\

\bottomrule
\end{tabular}
\end{sc}
\end{tiny}
\caption{\small {\bf BachProp on other datasets.} See \autoref{table:results} for description of labels.}
\end{table}

As a reference for future comparisons, we report here the results of BachProp trained on more complex datasets.
In Table \ref{table:results2}, we observe that for homogeneous corpora with many examples of similar structures (Chorales, Nottingham), BachProp can predict notes with higher accuracies than for more heterogeneous data sets (John Sankey, String Quartets).

We encourage readers to listen to the examples provided on the accompanying \href{https://sites.google.com/view/bachprop}{webpage} \cite{media} to convince themselves of the ability of BachProp and its variants to generate unique and heterogeneous new music scores. 

\section{Conclusion}

In this paper, we presented BachProp, an algorithm for general automated music composition. 
Our main contributions are (1) a note-sequence based representation of music with minimal distortion of the rhythm for training neural network models, (2) a network architecture that learns to generate pleasant music in this representation and (3) a set of metrics to compare generative models that operate on different representations of music. 

BachProp can be used both for automated and interactive music composition. Indeed, adding a human composer in the composition process is straightforward and BachProp can then be used as a computer aided music composition algorithm. For example, the human composer could use BachProp to suggest possible continuations and select among them. With such algorithms, the authors foresee that music composition can be brought to a wider audience, therefore allowing untrained humans to compose their own pieces of music. 


\begin{acknowledgments}
This research was partially supported by the Swiss National Science Foundation (Grant 200020\_165538) and the Laboratory of Computational Neuroscience (EPFL-LCN).
\end{acknowledgments} 

\bibliography{BachProp}

\begin{thebibliography}{10}
\providecommand{\url}[1]{#1}
\csname url@samestyle\endcsname
\providecommand{\newblock}{\relax}
\providecommand{\bibinfo}[2]{#2}
\providecommand{\BIBentrySTDinterwordspacing}{\spaceskip=0pt\relax}
\providecommand{\BIBentryALTinterwordstretchfactor}{4}
\providecommand{\BIBentryALTinterwordspacing}{\spaceskip=\fontdimen2\font plus
\BIBentryALTinterwordstretchfactor\fontdimen3\font minus
  \fontdimen4\font\relax}
\providecommand{\BIBforeignlanguage}[2]{{%
\expandafter\ifx\csname l@#1\endcsname\relax
\typeout{** WARNING: IEEEtran.bst: No hyphenation pattern has been}%
\typeout{** loaded for the language `#1'. Using the pattern for}%
\typeout{** the default language instead.}%
\else
\language=\csname l@#1\endcsname
\fi
#2}}
\providecommand{\BIBdecl}{\relax}
\BIBdecl

\bibitem{colton2012computational}
S.~Colton, G.~A. Wiggins \emph{et~al.}, ``Computational creativity: The final
  frontier?'' in \emph{ECAI}, vol.~12, 2012, pp. 21--26.

\bibitem{mordvintsev2015inceptionism}
A.~Mordvintsev, C.~Olah, and M.~Tyka, ``Inceptionism: Going deeper into neural
  networks,'' \emph{Google Research Blog. Retrieved June}, vol.~20, no.~14,
  p.~5, 2015.

\bibitem{gatys2016image}
L.~A. Gatys, A.~S. Ecker, and M.~Bethge, ``Image style transfer using
  convolutional neural networks,'' in \emph{IEEE Conference on Computer Vision
  and Pattern Recognition}.\hskip 1em plus 0.5em minus 0.4em\relax IEEE, 2016,
  pp. 2414--2423.

\bibitem{sturm2016music}
B.~L. Sturm, J.~F. Santos, O.~Ben-Tal, and I.~Korshunova, ``Music transcription
  modelling and composition using deep learning,'' in \emph{1st Conference on
  Computer Simulation of Musical Creativity}, 2016.

\bibitem{colombo2017deep}
F.~Colombo, A.~Seeholzer, and W.~Gerstner, ``Deep artificial composer: A
  creative neural network model for automated melody generation,'' in
  \emph{International Conference on Evolutionary and Biologically Inspired
  Music and Art}.\hskip 1em plus 0.5em minus 0.4em\relax Springer, 2017, pp.
  81--96.

\bibitem{hadjeres2016deepbach}
G.~Hadjeres, F.~Pachet, and F.~Nielsen, ``{D}eep{B}ach: a steerable model for
  {B}ach chorales generation,'' in \emph{34th International Conference on
  Machine Learning}, vol.~70.\hskip 1em plus 0.5em minus 0.4em\relax JMLR,
  2017, pp. 1362--1371.

\bibitem{lovelace1843notes}
A.~Lovelace, ``Notes on {L}. {M}enabrea's 'sketch of the analytical engine
  invented by {C}harles {B}abbage, esq.','' \emph{Taylor's Scientific Memoirs},
  1843.

\bibitem{fernandez2013ai}
J.~D. Fern{\'a}ndez and F.~Vico, ``{AI} methods in algorithmic composition: A
  comprehensive survey,'' \emph{Journal of Artificial Intelligence Research},
  vol.~48, pp. 513--582, 2013.

\bibitem{todd1989connectionist}
P.~M. Todd, ``A connectionist approach to algorithmic composition,''
  \emph{Computer Music Journal}, vol.~13, no.~4, pp. 27--43, 1989.

\bibitem{hochreiter1997long}
S.~Hochreiter and J.~Schmidhuber, ``Long short-term memory,'' \emph{Neural
  computation}, vol.~9, no.~8, pp. 1735--1780, 1997.

\bibitem{eck2002finding}
D.~Eck and J.~Schmidhuber, ``Finding temporal structure in music: Blues
  improvisation with {LSTM} recurrent networks,'' in \emph{12th IEEE Workshop
  on Neural Networks for Signal Processing}.\hskip 1em plus 0.5em minus
  0.4em\relax IEEE, 2002, pp. 747--756.

\bibitem{BL2012}
N.~Boulanger{-}Lewandowski, Y.~Bengio, and P.~Vincent, ``Modeling temporal
  dependencies in high-dimensional sequences: Application to polyphonic music
  generation and transcription,'' in \emph{29th International Conference on
  Machine Learning}, 2012.

\bibitem{Lattner2018}
S.~Lattner, M.~Grachten, and G.~Widmer, ``Imposing higher-level structure in
  polyphonic music generation using convolutional restricted {B}oltzmann
  machines and constraints,'' \emph{Journal of Creative Music Systems}, vol.~2,
  no.~1, 2018.

\bibitem{bachbotismir}
F.~Liang, M.~Gotham, M.~Johnson, and J.~Shotton, ``Automatic stylistic
  composition of {B}ach chorales with deep {LSTM},'' in \emph{18th
  International Society for Music Information Retrieval Conference}, 2017.

\bibitem{magenta2016polyphonyrnn}
{Magenta Team Google Brain}, ``{Polyphony RNN, revision ca73164},''
  \url{https://github.com/tensorflow/magenta/tree/master/magenta/models/polyphony_rnn},
  2016.

\bibitem{BachChorales}
``{J.S.}~{B}ach chorales,'' \url{http://web.mit.edu/music21/}, accessed:
  2019-04-01.

\bibitem{JohnSankey}
``{B}ach {MIDI} sequences by {J}ohn {S}ankey,''
  \url{http://www.jsbach.net/midi/midi_johnsankey.html}, accessed: 2019-04-01.

\bibitem{StringQuartets}
``String quartets by {M}ozart and {H}aydn,''
  \url{http://www.stringquartets.org}, accessed: 2019-04-01.

\bibitem{Theis2015}
L.~{Theis}, A.~{van den Oord}, and M.~{Bethge}, ``{A note on the evaluation of
  generative models},'' \emph{ArXiv:1511.01844}, p. arXiv:1511.01844, 2015.

\bibitem{media}
``{B}ach{P}rop media webpage,'' \url{https://sites.google.com/view/bachprop},
  accessed: 2019-04-01.

\bibitem{colombo2018bachprop}
F.~Colombo and W.~Gerstner, ``{B}ach{P}rop: Learning to compose music in
  multiple styles,'' \emph{arXiv preprint arXiv:1802.05162}, 2018.

\bibitem{colombo2018learning}
F.~Colombo, J.~Brea, and W.~Gerstner, ``Learning to generate music with
  {B}ach{P}rop,'' \emph{arXiv preprint arXiv:1812.06669}, 2018.

\bibitem{Sutskever2011}
I.~Sutskever, J.~Martens, and G.~Hinton, ``Generating text with recurrent
  neural networks,'' in \emph{28th International Conference on Machine
  Learning}, 2011, pp. 1017--1024.

\bibitem{Graves2013}
A.~{Graves}, ``{Generating Sequences With Recurrent Neural Networks},''
  \emph{ArXiv:1308.0850}, 2013.

\bibitem{Mikolov2012}
T.~Mikolov, ``Statistical language models based on neural networks,'' Ph.D.
  dissertation, 2012.

\bibitem{Oore2018}
S.~Oore, I.~Simon, S.~Dieleman, and D.~Eck, ``Learning to create piano
  performances,'' \emph{NIPS 2017 Workshop on Machine Learning for Creativity
  and Design}, 2017.

\bibitem{vandenOord2016}
A.~{van den Oord}, N.~{Kalchbrenner}, and K.~{Kavukcuoglu}, ``{Pixel Recurrent
  Neural Networks},'' \emph{ArXiv:1601.06759}, 2016.

\bibitem{chung2014empirical}
J.~Chung, C.~Gulcehre, K.~Cho, and Y.~Bengio, ``Empirical evaluation of gated
  recurrent neural networks on sequence modeling,'' \emph{arXiv preprint
  arXiv:1412.3555}, 2014.

\bibitem{karpathy2016unreasonable}
A.~Karpathy, ``The unreasonable effectiveness of recurrent neural networks,''
  \url{http://karpathy.github.io/2015/05/21/rnn-effectiveness}, 2016.

\bibitem{chollet2015keras}
F.~Chollet \emph{et~al.}, ``Keras,'' \url{https://github.com/fchollet/keras},
  2015.

\bibitem{Nottingham}
``Nottingham data set of folk songs,''
  \url{http://www-etud.iro.umontreal.ca/~boulanni/icml2012}, accessed:
  2019-04-01.

\bibitem{hadjeresmaster}
``{D}eep{B}ach, revision f069695,''
  \url{https://github.com/Ghadjeres/DeepBach}, accessed: 2019-04-01.

\bibitem{kingma2014adam}
D.~P. Kingma and J.~Ba, ``Adam: A method for stochastic optimization,''
  \emph{arXiv preprint arXiv:1412.6980}, 2014.

\end{thebibliography}

\end{document}

%
%